\documentstyle[12pt,epsf]{article}

 \catcode`\@=11
\def\Bbb{\ifmmode\let\next\Bbb@\else
 \def\next{\errmessage{Use \string\Bbb\space only in math mode}}\fi\next}
\def\Bbb@#1{{\Bbb@@{#1}}}
\def\Bbb@@#1{\fam\msbfam#1}

\if@twoside
\else
\fi
\ifcase \@ptsize
\or 
\or 
\fi

\def\@citex[#1]#2{%
\if@filesw \immediate \write \@auxout {\string \citation {#2}}\fi
\@tempcntb\m@ne \let\@h@ld\relax \def\@citea{}%
\@cite{%
  \@for \@citeb:=#2\do {%
    \@ifundefined {b@\@citeb}%
      {\@h@ld\@citea\@tempcntb\m@ne{\bf ?}%
      \@warning {Citation `\@citeb ' on page \thepage \space undefined}}%
      {\@tempcnta\@tempcntb \advance\@tempcnta\@ne%
      \@tempcntb\number\csname b@\@citeb \endcsname \relax%
      \ifnum\@tempcnta=\@tempcntb 
        \ifx\@h@ld\relax%
          \edef \@h@ld{\@citea\csname b@\@citeb\endcsname}%
        \else%
          \edef\@h@ld{\ifmmode{-}\else--\fi\csname b@\@citeb\endcsname}%
        \fi%
      \else
        \@h@ld\@citea\csname b@\@citeb \endcsname%
        \let\@h@ld\relax%
      \fi}%
    \def\@citea{,\penalty\@highpenalty\,}%
  }\@h@ld
}{#1}}
\@addtoreset{equation}{section}
\def\section{\@startsection {section}{1}{\z@}{-3.5ex plus -1ex minus
 -.2ex}{2.3ex plus .2ex}{\large\bf\centering}}
\def\subsection{\@startsection{subsection}{2}{\z@}{-3.25ex plus -1ex minus
 -.2ex}{1.5ex plus .2ex}{\sc}}
\gdef\@publabel{\hfil}
\gdef\@pubdate{\null}
\gdef\@pubnumber{\null}
\gdef\@author{\null}
\gdef\@title{\null}
\gdef\@abstract{\null}
\long\def\pubdate#1{\gdef\@pubdate{#1}}
\long\def\pubnumber#1{\gdef\@pubnumber{#1}}
\long\def\publabel#1{\gdef\@publabel{#1}}
\long\def\author#1{\gdef\@author{#1}}
\long\def\title#1{\gdef\@title{#1}}
\long\def\abstract#1{\gdef\@abstract{#1}}
\def\titlerelax{
\let\maketitle\relax
\let\settitleparameters\relax
\let\consolidatetitle\relax
\let\inittitlepage\relax
\let\finishtitlepage\relax
\let\titlepagecontents\relax
\let\multithanks\relax
\let\titlebaselines\relax
\let\@makepub\relax
\let\@maketitle\relax
\let\@makeauthor\relax
\let\@makeabstract\relax
\let\@maketitlenote\relax
\let\thanks\relax
\let\titlerelax\relax}
\def\titleclean
{\gdef\@titlenote{}
\gdef\@abstract{}
\gdef\@author{}
\gdef\@title{}
\gdef\@pubdate{}\gdef\@pubnumber{}\gdef\@publabel{}
\gdef\@dpublabel{}
}

\def\@makepub{\vbox to \z@{\hbox to \textwidth{\hfill
\@publabel \hfill
\llap{\parbox[t]{0.25\textwidth}{\raggedleft\@pubnumber}}}%
\vss}}
\def\@maketitle{\vskip 60pt \begin{center}
 {\LARGE \@title \par}
 \end{center}}
\def\@makeauthor{{%
\def\and{\smallskip {\normalsize \rm and\smallskip }}
\def\And{\medskip {\normalsize \rm and\\}\medskip}
\long\def\address##1{{\def\and{\\and\\}\medskip
				{\small \it \\##1\\}
}}
{\centering
 \vskip 3em
 \large \lineskip .75em
 \@author}
 \par}}
\def\@makedate{\vskip 1.5em
 {\raggedright \small \noindent\@pubdate \par}}
\def\@makeabstract{\vskip 1.5em
{\small
\begin{center}
{\bf ABSTRACT\vspace{-.5em}\vspace{0pt}}
\end{center}
\quotation \@abstract \endquotation}}
\def\maketitle{\titlepage
\let\footnotesize\small \setcounter{page}{0}
\@makepub
\vfil
\@maketitle
\@makeauthor
\vfil
\@makeabstract
\@thanks
\vfil
\@makedate
\if@restonecol\twocolumn \else \eject \fi
\titlerelax \titleclean
\setcounter{footnote}{0}
}

\catcode`\@=12

\begin{document}
\bibliographystyle{npb}

\let\b=\beta
\def\blank#1{}

\def\cdd{{\cdot}}
\def\cev#1{\langle #1 \vert}
\def\cH{{\cal H}}
\def\comm#1#2{\bigl [ #1 , #2 \bigl ] }
\def\compact{ reductive}
\def\cont{\nonumber\\*&&\mbox{}}
\def\cO{{\cal O}}
\def\cul #1,#2,#3,#4,#5,#6.{\left\{ \matrix{#1&#2&#3\cr #4&#5&#6} \right\}}

\def\dz{Dz}
\def\dz{\hbox{$d\kern-1.1ex{\raise 3.5pt\hbox{$-$}}\!\!z$}}
\def\dz{ \frac{d\!z}{2\pi i}}
\def\en{\end{equation}}
\def\enn{\end{eqnarray}}
\def\eq{\begin{equation}}
\def\eqq{\begin{eqnarray}}



\def\half#1{\frac {#1}{2}}

\def\ip#1#2{\langle #1,#2\rangle}


\def\k{k}


\def\Mf#1{{M{}^{{}_{#1}}}}
\def\mno{{\textstyle {\circ\atop\circ}}}
\def\mod#1{\vert #1 \vert}

\def\Nf#1{{N{}^{{}_{#1}}}}
\def\ni{\noindent}
\def\no{{\textstyle {\times\atop\times}}}
\def\no:#1:{\mno#1\mno}
\def\nox{{\scriptstyle{\times \atop \times}}}


\let\p=\phi
\def\posdef{ positive-definite}
\def\posdefness{ positive-definiteness}
\def\Qf#1{{Q{}^{{}_{#1}}}}
\def\Qstar{\mathop{\no:QQ:}\nolimits}

\def\reductive#1#2{#1}


\def\tr{\mathop{\rm tr}\nolimits}
\def\Tr{\mathop{\rm Tr}\nolimits}







\def\vec#1{\vert #1 \rangle}
\def\vac{\vec 0}

\def\wan{$\WA_n$ }
\def\Wb{\bar W}
\def\Wf#1{{W{}^{{}_{#1}}}}
\def\wbn{$\WB_n$ }
\def\WA{\mathop{\it WA}\nolimits}
\def\WB{\mathop{\it WB}\nolimits}
\def\WBC{\mathop{\it WBC}\nolimits}
\def\WD{\mathop{\it WD}\nolimits}
\def\WG{\mathop{\it WG}\nolimits}



\def\zz#1{(z-z')^{#1}}

\openup 1\jot

\pubnumber{ DTP 96-37}
\pubdate{September 1996}

\title{Classical backgrounds and scattering for affine Toda theory on a half-line}

\author{
P. BOWCOCK\thanks{ Email \tt
Peter.Bowcock@durham.ac.uk}
\address{Dept. of Mathematical Sciences,
University of Durham,  
Durham, DH1 3LE, U.K.}
}

\abstract{We find classical solutions to the simply-laced affine Toda equations which satisfy integrable boundary conditions using solitons which are 
analytically continued from imaginary coupling theories. Both static 
`vacuum' configurations and time-dependent perturbations about them
which correspond respectively to classical vacua  and particle
scattering solutions are considered. 
A large class of classical scattering matrices are calculated and 
found to satisfy the reflection bootstrap equation. }
\maketitle

\section{Introduction}

Integrable models in two-dimensions possess remarkable features. In particular
the S-matrix for such a theory factorises into products of 
two-particle scattering matrices $S^{ab}_{cd}(\theta)$ (which gives the matrix
element for the process $a+b \to c+d$) \cite{ZZ}. Under reasonable physical assumptions,
it is possible to show that $S$ must satisfy a number of rather simple 
algebraic equations. Solutions have been found to these equations for all
affine Toda theories and these
are postulated to give a non-perturbative expression for the scattering 
matrix of the theory \cite{BCDS,DGZ,CM}. 

These ideas have been extended to include
integrable theories on a half-line \cite{C,FK1,GZ}. 
In this case it is necessary to introduce
another matrix $K$ which describes the reflection of particles off the boundary.
Again it is generally believed that all scattering amplitudes factorise 
into products of $S$ and $K$. In addition the algebraic equations which were
solved by $S$ alone previously can be modified to include $K$. Recently
a number of solutions to these equations for $S$ and $K$ have been found
in the context of affine Toda theories \cite{CDR,CDRS,FK1,FK2,Ki,S}. 
However, many of these solutions make no reference to the 
boundary conditions which presumably need to be imposed to make sense of 
the theory. (The exceptions to this are \cite{Ki} whose perturbative analysis
is tied to Neumann boundary conditions and \cite{CDR,CDRS}).
In fact, integrability places 
severe constraints on the boundary conditions one can impose; for
simply-laced affine Toda theories there are only a finite number of 
possibilities \cite{CDR,CDRS,BCDR}.

Fortunately, there seems to be a straightforward way of analysing which 
solutions for $S$ and $K$ correspond to which boundary conditions . Whilst
$S$ is known to tend to unity in the classical limit, it seems that in  
general the reflection factor $K$ does not. Thus if we know the 
classical reflection factors for a particular boundary condition,
it should be possible to identify the corresponding quantum reflection
factor by considering its classical limit \cite{CDR}.

The aim of this paper is to make some progress towards calculating 
classical scattering for Toda theories on a half-line. This problem
naturally divides into two parts.
{}First it is necessary to find the `vacuum configuration' or lowest 
energy solution (which is presumably static) which satisfies a particular
boundary condition. Then one solves the linear equations for infinitesimal
perturbations about this solution still ensuring that the boundary 
conditions are satisfied. Far from the wall the solution consists
of a superposition of incoming and outgoing waves. The relative phase 
between the waves is interpreted as the classical limit of the reflection
factor $K$. Some examples of calculations along these lines are to be found
in \cite{CDR} where particular cases within $a_2^{(1)},d_5^{(1)}$ and 
$a_1^{(1)}$ were considered.

The paper will be divided into five sections. In the next section we review
affine Toda theories, and in particular the boundary conditions which can
be imposed which are consistent with the (classical) integrability of the 
theory \cite{BCDR,BCR}. 
It is then shown that the boundary conditions combine in a 
particularly neat way with the equations of motion to give  linear 
equations that the tau functions must satisfy at the boundary. 
This is the key result of the paper which enables us to solve for the 
classical scattering solutions. 

In the third section we consider static solutions to the boundary conditions
for the special case of $a_n^{(1)}$ Toda theories.
For these theories a large number of tau functions can be constructed 
explicitly by analytically continuing the multi-soliton solutions of the 
imaginary coupling theory  \cite{H}. 
Whilst it can be shown that these solutions are
singular on the whole line, it is possible that all the singularities can
be placed `behind' the boundary for a theory on the half-line. By substituting
this family of tau-functions into the equation derived in the previous section
we obtain a large number of possible solutions. The correspondence with 
particular boundary conditions is discussed.

In section four, whilst remaining within the context of $a_n^{(1)}$ Toda
theories, the analysis is extended 
to cope with the scattering perturbations
around vacuum solutions. In this way we are able to derive an expression for
the classical reflection matrix. It is shown that this satisfies the 
classical reflection bootstrap equation. 

We conclude with comments on our results and directions for future research.

\section{Classical affine Toda theory on a half-line}

In this section we briefly review some of the features of affine Toda theory
on a half line that we shall need later on to establish notation.

To each affine Kac-Moody algebra $\hat g$ we can associate an affine Toda 
theory \cite{MOP}. For simplicity in this paper we shall restrict ourselves to 
simply-laced algebras, and for the most part to $a_n^{(1)}$ which are in some
senses the simplest cases of all.
The equations of motion for the affine Toda theory associated to $\hat g$ are
given by 
\eq
\partial_{\mu} \partial^{\mu}\phi+{m\over \beta}\sum_{i=0}^r n_i \alpha^i
e^{\beta \alpha^i \cdot \phi}=0.
\label{eqm}
\en
Here we have used the notation that $\alpha^i$ for $i=1$ to $r$ are the 
simple roots of $g$, the finite Lie algebra associated to $\hat g$, and 
$\alpha^0=-\psi$ where $\psi$ is the highest root of $g$ 
(for untwisted algebras).  Also we have
defined $r$ to be the rank of $g$ and introduced $m$ the mass parameter, and 
$\beta$ the coupling constant of the theory. In this paper we shall consider
theories for which $\beta$ is real, and henceforth we set $\beta=m=1$.
The constants $n_i$ are given by the equation
\eq
\alpha_0+\sum_{i=1}^r n_i \alpha^i=0
\en
and are sometimes referred to as  `marks'. 

When the classical theory is considered
on the half line $-\infty\le x\leq 0$, we need to supplement the equations
of motion by boundary conditions. In this paper we shall restrict ourselves
to conditions of the form
\eq
\partial_x \phi=F(\phi)
\en
although more general possibilities can be considered \cite{NW,BCR}. 
An arbitrary choice of $F(\phi)$ will generically break the integrability of 
the system. It turns out that for 
simply-laced theories (with the exception of sinh-Gordon or $A^{(1)}_1$
affine Toda theory \cite{M,T}) 
there are only a finite number of choices for $F$ which preserve 
integrability \cite{CDR,CDRS,BCDR}.
These can be summarised as follows:

\noindent
(1) Either
\eq
\partial_x \phi= 0;
\label{eq.Neumann}
\en
these are `free' or Neumann boundary conditions

\noindent
(2) Or
\eq
\partial_x \phi= \sum_{i=0}^r A_i \alpha^i \sqrt{ n_i} e^{\alpha^i \cdot \phi/2}
\label{eq.non.Neumann}
\en
where the $A_i=\pm 1$.

The first of these two possibilities is fairly easy to analyse completely at 
a classical level. The classical energy functional on the half-line is 

\eq
E=\int_{-\infty}^0 dx \left ({1\over 2}(\partial_x \phi)^2+
{1\over 2}(\partial_t \phi)^2+\sum_{i=0}^r n_i (e^{\alpha^i \cdot \phi} -1)
\right ).
\label{eq.bulk.energy}
\en
This is non-negative and the lowest energy configuration is clearly $\phi=0$
which also satisfies the Neumann boundary conditions. If we now take the field
$\phi(x,t)=\epsilon(x,t)$ to be some infinitesimal perturbation to this 
vacuum configuration we see that in a linear approximation we have 
\eq
\partial_{\mu} \partial^{\mu} \epsilon +M \epsilon =0
\label{eq.wave}
\en 
where $M=\sum_{i=0}^r n_i \alpha^i \otimes \alpha^i$ is the mass matrix. The 
eigenvectors $\rho_a$, $a=1$ to $r$, of $M$ are in one-to-one 
correspondence with the 
fundamental representations of the Lie algebra $g$ and are interpreted as 
the basic particle-like excitations of the theory. The eigenvalue of $\rho_a$
is given by $\lambda_a=m_a^2$
where $m_a$ is the mass of the 
corresponding particle. It is a remarkable fact that the set of masses
form the lowest eigenvalue eigenvector of the Cartan 
matrix of $g$ \cite{BCDS,F,FLO}. 
In terms of $\rho$ the basic solutions of \ref{eq.wave} are given by
\eq
\phi(x,t)=\epsilon(x,t)=\rho_a e^{iEt}(e^{ipx}+K_a e^{-ipx})
\label{eq.sol}
\en 
where $E^2-p^2=m_a^2$. This is a superposition of left- and right-moving 
waves. The classical reflection factor is given by the phase factor $K_a$
which can be determined by substituting the solution (\ref{eq.sol})  
into the boundary conditions (\ref{eq.Neumann}). This yields the solution 
$K_a=1$. Classically, at any rate, this boundary condition seems rather 
uninteresting. There has been considerable progress in understanding 
the quantum case in \cite{Ki}, where the semi-classical reflection 
factor has been calculated using perturbative techniques.

In this paper we shall be interested in 
boundary conditions of the form (\ref{eq.non.Neumann}). It turns out to 
be particularly convenient to analyse this case in the language of {\it 
tau-functions} \cite{H}. 
These are introduced as a particular parametrisation of 
the function $\phi$;
\eq
\phi=-\sum_{i=0}^r  \alpha^i ln(\tau_i)
\label{eq.def.tau}
\en
Note that we have introduced $r+1$ functions $\tau_i$ to describe the $r$-component
field $\phi$ and this is reflected in the freedom to scale 
all the tau-functions $\tau_i\to f(x,t)\tau_i$
without affecting the value of $\phi$. This unphysical degree of freedom is 
fixed by demanding that $\tau_i$ satisfies the following particular
form of the equations of motion
\eq
\ddot{\tau_i} \tau_i -\dot{\tau_i}^2-\tau_i''\tau_i +(\tau_i')^2
=\left (\prod_{j=0}^r \tau_j^{I_{ij}}-\tau_i^2\right )n_i,
\label{eq.tau.eqm}
\en
where $I_{ij}$ is the incidence matrix of $g$ given by
\eq 
I_{ij}=2\delta_{ij}-\alpha^i \cdot \alpha^j
\en
Essentially, $I_{ij}$ takes the value one if the nodes corresponding to $i$ and
$j$ are connected on the extended Dynkin diagram, and vanishes otherwise.
Substituting the form (\ref{eq.def.tau}) into the boundary conditions
(\ref{eq.non.Neumann}), and taking the inner product with the fundamental 
weight $\lambda_i$, we discover that the boundary conditions can be 
written in terms of the tau-functions as
\eq
{\tau_i' \over \tau_i}+\sqrt{n_i} A_i e^{\alpha^i \cdot \phi /2}=
n_i C
\en
where
\eq 
C=({\tau_0'\over \tau_0} +A_0 e^{\alpha^0 \cdot \phi /2}).
\en
From this we see that
\eq
n_i \prod_j \tau_j^{I_{ij}}=\tau_i^2 (n_i C- {\tau_i' \over \tau_i})^2
\en
Substituting this into the equations of motion (\ref{eq.tau.eqm}) we find
\eq 
\ddot{\tau}_i-{(\dot{\tau_i})^2 \over \tau_i}-\tau_i''+2 n_i C \tau_i'
-(n_i^2 C^2 -n_i)\tau_i =0.
\label{eq.master}
\en
This equation is the main result of this section. Immediately we see that 
it has two attractive features. Firstly the equations for $\tau_i$ have 
decoupled so that each equation only involves $\tau_i$ for some $i$.
Secondly, if we assume that the solution is static then the first
two terms vanish and we are left with the linear equation
\eq
\tau_i''-2 n_i C \tau_i' +(n_i^2 C^2-n_i)\tau_i=0
\label{eq.static.master}
\en
At this point we perhaps should remind the reader that this equation 
is deduced from the equations of motion and the boundary conditions, and 
that the latter are only valid at the boundary $x=0$. Thus the above 
equation cannot be solved as a differential equation in $x$ since it is 
only valid at the boundary, but it will prove a valuable tool in 
determining the subset of solutions of the equations of motion which 
satisfy one of the boundary conditions (\ref{eq.non.Neumann}). One of 
the drawbacks of the equation (\ref{eq.static.master}) is that we
have essentially squared the boundary conditions, so that we have lost the
information contained in the $A_i$. This we will have to recover by 
explicitly examining the proposed solutions at the boundary

Perhaps even more surprisingly we can use a similar equation in the
case of time-dependent perturbations to a static vacuum solution, i.e. in
the situation where 
\eq
\tau_i(x,t)=(\tau_i(x))_{vac}+\epsilon_i(x,t)
\en
where $\epsilon_i(x,t)$ is some infinitesimal perturbation to the vacuum
solution $(\tau_i)_{vac}$. In this case, the important point is that 
$\dot{\tau}_i=O(\epsilon)$, so discarding terms of $O(\epsilon^2)$ in 
(\ref{eq.master}) we again arrive at the linear equation
\eq 
\ddot{\tau}_i-\tau_i''+2 n_i C \tau_i'
-(n_i^2 C^2 -n_i)\tau_i =0
\label{eq.td.master}
\en
We shall use this equation in section four to solve for classical scattering
about the vacuum.

One may be concerned that one has introduced a spurious constant $C$ in
equation (\ref{eq.static.master}). Remarkably $C$ has a 
physical interpretation as proportional to the 
energy of the solution corresponding to
$\tau_i(x,t)$ on the half-line. The energy on the half-line for boundary
conditions of type (\ref{eq.non.Neumann}) is given by 
\eq
E=E_{bulk}-\sum_{i=0}^r 2 A_i \sqrt{n_i} e^{\alpha^i \cdot \phi/2}|_{x=0}
\en
where $E_{bulk}$ is the bulk energy given by (\ref{eq.bulk.energy}) and
the second term is the contribution to the energy from the boundary 
compatible with the boundary conditions (\ref{eq.non.Neumann}).
It has been shown in \cite{OTU} that for soliton solutions 
the energy density is a total derivative, so that the bulk energy can 
be written as a boundary contribution which in terms of tau-functions is
\eq
E_{bulk}=\left [ -2 \sum_{i=0}^r {\tau_i'\over \tau_i} \right ]^0_{x=-\infty}.
\label{eq.der.energy}
\en
In the sections that follow we shall be using tau-functions that tend
to a constant as $x\to -\infty$ so the contribution to (\ref{eq.der.energy})
come only from $x=0$. Adding the contribution from the bulk and the boundary
we find that
\eqq
E&=&-2\sum_{i=0}^r \left ( {\tau_i' \over \tau_i} + A_i \sqrt{n_i} e^{\alpha^i
\cdot \phi /2} \right )\\
&=&-2 \sum_{i=0}^r n_i C= -2 h C
\label{eq.total.energy}
\enn
where $h$ is the Coxeter number associated to $g$. Thus, although
initially it may have seemed that the appearance of $C$ in the equation (\ref{eq.static.master}) was a drawback, it turns out that it is an added
bonus, giving the energy of the solution that we are 
considering. This is important in trying to determine
the `vacuum' for a given boundary condition, since the vacuum
is defined to be the static solution
with the lowest energy which satisfies the boundary condition.

\section{Static vacuum configurations for $a_n^{(1)}$ Toda theories from
analytically continued solitons}

In this section we shall confine ourselves to affine Toda theories
based on the algebra $a_n^{(1)}$. We have seen in the previous section that
the requirement that a solution satisfies one of the integrable boundary
conditions of the form (\ref{eq.non.Neumann}) can be neatly expressed in
terms of tau-functions. For the $a_n^{(1)}$  we have a particularly
rich source of such tau-functions which can be constructed by analytically
continuing multi-soliton solutions of the imaginary coupling theory \cite{H}.

At this point let us briefly review the nature of these solutions in the 
imaginary coupling theory.
Let us reintroduce the coupling constant, so that we write
\eq
\beta \phi=-\sum_{i=0}^r \alpha^i ln(\tau_i)
\label{eq.def.tau2}
\en
The tau-functions for an $N$-soliton solutions can be compactly written as
\eq
\tau_j(x,t)=\sum_{\mu_1=0}^1\dots\sum_{\mu_N=0}^1 {\rm exp}
\left ( \sum_{p=1}^N \mu_p \omega^{a_p j} \Phi_p +\sum_{1\leq p\le q\leq N}
\mu_p\mu_q ln A^{(a_p a_q)}\right ).
\label{eq.multi.sol}
\en
Here we have introduced the notation $\Phi_p=\sigma_p(x-v_p t)+\xi_p$
where $v_p$ is the velocity of the $p$-th soliton, $\xi_p$ is a complex
parameter whose real part and imaginary part are related respectively to
the position and the topological charge of the $p$-th soliton. Also 
$\sigma_p$ and $v_p$ are related by the mass-shell condition
\eq
\sigma_p^2(1-v_p^2)=m_{a_p}^2
\label{eq.mass.shell}
\en
where $a_p$ labels the species of soliton and $m_{a_p}=2\sin({{\pi a_p}\over {n+1}})$. We define $\omega=exp(2\pi i /(n+1))$.
The variables $\sigma_p$ and $v_p$ are often conveniently 
parametrised in terms of the two-dimensional rapidity $\theta_p$ by putting
\eqq
\sigma_p &=& m_{a_p} \cosh(\theta_p)\\
\sigma_p v_p &=& m_{a_p} \sinh(\theta_p)
\label{eq.rap.one}
\enn
The constants $A^{(a_p a_q)}$ describe the interaction between the $p$-th and
$q$-th solitons and are given as
\eqq
A^{(a_p a_q)}(\Theta)
&=&-{{(\sigma_p-\sigma_q)^2-(\sigma_p v_p-\sigma_q v_q)^2-4 \sin^2
{\pi \over {n+1}}(a_p-a_q)}\over {(\sigma_p+\sigma_q)^2-
(\sigma_p v_p+\sigma_q v_q)^2-4 \sin^2
{\pi \over {n+1}}(a_p+a_q)}}\\
&=& {{\sin({\Theta \over 2i}+{\pi(a_p-a_q)\over {2(n+1)}})
\sin({\Theta \over 2i}-{\pi(a_p-a_q)\over {2(n+1)}})}\over 
{\sin({\Theta \over 2i}+{\pi(a_p+a_q)\over {2(n+1)}})
\sin({\Theta \over 2i}-{\pi(a_p+a_q)\over {2(n+1)}})}}
\label{eq.inter.def}
\enn
where $\Theta=\theta_p-\theta_q$. In this section we shall only be 
interested in static solitons so we shall take $\theta_p=0$ or equivalently
$v_p=0$, $\sigma_p=m_{a_p}$. 

For single solitons the expression (\ref{eq.multi.sol})
reduces to 
\eq
\tau_j=1+\omega^{j a} e^{\Phi_p}
\en
The topological charge of such a soliton of species $a$ which is defined
by 
\eq
\lim_{x\to \infty} (\phi(x,t)-\phi(-x,t))
\en
can be shown to lie in the fundamental representation with highest
weight $\lambda_{a}$ where $\lambda_a\cdot \alpha^b= \delta_{a}^b$. 
It is therefore natural to associate each species of soliton with nodes on the 
Dynkin diagram of $a_n$. By analogy with the representation theory of 
$a_n$ we refer to solitons of type $a$ and type $\bar{a}=n+1-a$ as conjugate. 
We shall (ab)use the notation $\bar{p}$ to denote a soliton whose type is
congugate to that of soliton $p$.

Now let us specialise to the case of interest, namely static solitons.
Examining the expression (\ref{eq.multi.sol}) carefully
we see that solitons obey a kind of Pauli-exclusion principle. We can
only construct multi-soliton solutions whose constituent solitons have 
either different velocity and/or different species. If we attempt to 
consider two solitons of the same species and velocity, the interaction
constant $A^{(aa)}(0)$ vanishes and we simply end
up with one constituent soliton of that velocity and species at some
different position. This places severe constraints on the possibilities
for static soliton configurations; since the velocities of each constituent
is identically zero, it follows that each constituent 
soliton must be of different species. Thus we can consider at most
an $n$-soliton solution where each constituent soliton corresponds to 
a different node on the Dynkin diagram of $a_n$. Indeed we believe that 
this is the most general static solution whose energy-density tends to zero
at spatial infinity.
The energy of a stationary soliton of type $a$ on the whole line
is simply given by
\eq
E=-{2\over \beta^2}(n+1) m_a
\label{eq.single.energy}
\en
Note that for imaginary $\beta$ this is positive, as we might expect. The 
energy of $N$ stationary solitons is simply given as the sum of energies
of the constituent solitons $E=\sum_p E_p$.  

The above discussion has all been in the context of imaginary coupling
Toda theory, where it is accepted that the field $\phi$ is allowed to 
be complex. We are interested in real coupling Toda theory where  
the field is $\phi$ is taken to be real. A little thought shows that for 
tau-functions of the type given in (\ref{eq.multi.sol}), reality of $\phi$
implies the reality of $\tau_j$. This can be ensured by insisting that 
each constituent soliton $p$ is paired with a congugate soliton $\bar{p}$
with the position/topological charge variables related by 
$\xi_p=(\xi_{\bar{p}})^*$. An exception to this rule occurs for $n$ odd,
where the soliton corresponding to the middle node of the dynkin
diagram, i.e. $a_p=(n+1)/2$, is unpaired (since it is its own conjugate), 
so we must take$\xi_p$ to be real in this case. Thus the reality
of the tau-functions closely corresponds to the representation theory
of $a_n$, where only the middle node corresponds to a real representation
and the other fundamental representations must be taken in conjugate 
pairs if we wish to restrict ourselves to real representations. Similar
remarks apply equally well to tau-functions associated with other 
simply-laced algebras

Finally, let us note that the energy $E$ for such solutions on the whole
line given in (\ref{eq.single.energy}) is negative for real $\beta$. This 
may seem surprising in view of the manifestly positive energy density of the 
theory (\ref{eq.bulk.energy}). This reflects the fact that all such 
solutions become singular somewhere on the real line, and this is perhaps
why they are not generally considered in the 
real coupling theories. However, in the present context the possibility
exists that all the singularities in $\phi$ lie in the `unphysical' region
$x>0$, so that they are perfectly acceptable physically.

\subsection{Two-soliton solutions}

As a pedagogical introduction to a more general solution, we begin by 
considering the possible static two-soliton solutions that satisfy the 
boundary conditions (\ref{eq.non.Neumann}). This solution and some of its
features have already been discussed in \cite{FU}. A two-soliton solution 
consisting of a soliton of type $a$ and its anti-soliton of type $\bar{a}$ have 
tau-functions of the form
\eq
\tau_i=1+2 d \cos(\chi+{2\pi i a\over n+1})e^{m_a x}+A^{(a\bar{a})} d^2 
e^{2 m_a x}
\label{eq.two.soliton}
\en
Here we have set $\chi= Im(\xi)$ and $d={\rm exp}(Re(\xi))$ in the 
previous notation. Also
we calculate $A^{(a\bar{a})}$ from setting $\Theta$ to zero in 
(\ref{eq.inter.def}) as 
\eq
A^{(a\bar{a})}=\cos^2({\pi a\over n+1})=1-{m_a^2 \over 4}.
\label{eq.appbar}
\en
In general multi-soliton tau-functions (\ref{eq.multi.sol}) can be split up 
into a sum of `charge' sectors by writing
\eq
\tau_j= \sum_k T_k \omega^{k j}.
\label{eq.charge.tau}
\en
The linearity of 
equation (\ref{eq.static.master}) and the fact that for $a_n$, we have that 
$n_i=1$ for all $i$ implies that
each of the $T^k$ separately satisfy the equation; 
\eq
(T_k)''-2 C (T_k)'+(C^2-1)T_k=0\;\;\;{\rm at}\;\;x=0
\label{eq.mod.stat}
\en
In the two-soliton case there are three charge sectors:$k=0,k=\pm a$. Obviously $T_a=
\bar{T}_{-a}$, so there are essentially two independent equations;
\eqq
(C^2-1)+((C-2 m_a)^2-1)A^{(a\bar{a})}d^2&=&0\\
((C-m_a)^2-1)d &=&0
\enn
The second of these equations implies that for a non-trivial solution
\eq
C_{\pm}=m_a\pm 1,
\en
and, substituting this into the first we find that
\eq
d={2\over (2\mp m_a)}.
\en

The energy of the two solutions
are given by $E_{\pm}=-2 h C_{\pm}=-2(n+1)(m_p\pm 1)$. Actually, it is clear that solutions should come in pairs. The reason is 
that if $\phi(x,t)$ is a solution to the equations of motion and the 
boundary conditions, then so is the parity reversed solution $\phi(-x,t)$,
since it satisfies boundary conditions of the form (\ref{eq.non.Neumann}) but
with $A_i\to -A_i$. Note that if we sum the energies of the two solutions,
the boundary contributions of each solution cancel, and we are left with
a bulk contribution of $\phi(x,t)$ and $\phi(-x,t)$ from $-\infty<x<0$ which
can be rewritten as simply the bulk energy of $\phi(x,t)$ on the whole line,
and indeed we find that
\eq
E_+ +E_- =-2(n+1)m_a 
\en 
which is the energy of two static solitons of type $a$.

Before we pick the lower energy 
solution as the vacuum we should be careful to consider whether the 
singularities of the corresponding solution for $\phi$ lie in the physical
region $x<0$ or the unphysical region $x>0$. Since we have established that
the two solutions are parity conjugate, and we know that the solution is 
singular at some point on the whole line, it follows that at least one
solution must be singular somewhere in the physical region. (It is possible
that the singularity lies at $x=0$ which is energetically allowed. We discuss
this case later). It turns out that the `good' solution whose singularities
lie in the region $x>0$ corresponds to the higher
energy solution $C_-$. For this solution the 
tau-function can be written
\eq
\tau_j={{(2+m_a)+4\cos(\chi+{{2\pi j a}\over n+1})e^{m_a x}+(2-m_a)e^{2 m_a x}}
\over 2+m_a}
\label{eq.tau.two}
\en
Singularities in $\phi$ occur when $\tau_j$ vanishes for some $j$. Solving the
equation $\tau_j=0$ yields
\eq
e^{m_a x}={{-\cos(\chi+{{2\pi j a}\over n+1})\pm\sqrt{\cos^2(\chi+{{2\pi j a}\over n+1})-\cos^2({\pi a\over n+1})}}\over {1-\sin({\pi a\over n+1})}}
\label{eq.solv.sing}
\en
For a real solution for $x$ this implies that 
\eqq
\cos^2(\chi+{{2\pi j a}\over n+1})&\geq&\cos^2({\pi a\over n+1})\\
\cos(\chi+{{2\pi j a}\over n+1})&\leq &0.
\enn
Combining these two conditions gives
\eq
-1\leq \cos(\chi+{{2\pi j a}\over n+1}) \leq -\cos({\pi a\over n+1})
\en
It is easy to check that for any choice of $\chi$ there will be some
$j$ for which this condition is satisfied, so that there will always be 
a singularity at some real value of $x$ for some $j$. 
To see where these singularities lie consider the right hand side of 
(\ref{eq.solv.sing}) where the square-root is taken with the minus sign,
since this is the smaller of the two solutions, and hence the more likely
to yield negative values of $x$.  A simple calculation shows that 
as a function of $\cos(\chi+{{2\pi j a}\over n+1})$ the right hand side is 
a monotonically increasing function, so its lowest value is attained when
$\cos(\chi+{{2\pi j a}\over n+1})=-1$. At this point the right hand side is
equal to one, corresponding to a singularity at $x=0$. For other values
of $\cos(\chi+{{2\pi j a}\over n+1})$, the right hand side will be greater than
one, and the singularities will be in the unphysical region $x>0$.

Having found a class of two soliton solutions which correspond to non-singular fields $\phi$, it only remains to ascertain precisely which boundary conditions
each solution satisfies. Substituting $x=0$ into the expression (\ref{eq.tau.two}), we find that the tau-function at this point can be
written as 
\eq
\tau_j={\cos^2({\chi\over 2}+{\pi j a\over n+1}) \over {1+\sin({\pi a\over n+1})}}
\en
Thus we can write 
\eq
e^{\alpha^j \cdot \phi} ={ {\cos^2({\chi\over 2}+{\pi (j+1) a)\over n+1})\cos^2({\chi\over 2}+{\pi (j-1) a \over n+1})}\over 
\cos^4({\chi\over 2}+{\pi j a\over n+1})}
\en
In the boundary conditions (\ref{eq.non.Neumann}), we have terms involving
${\rm exp}(\alpha^j \cdot \phi/2)$ which should be interpreted as the 
positive square root of the above quantity. The values of $A_i$ are
found to be given by
\eq
A_j=-{\rm sign}\left ({\cos({\chi\over 2}+{\pi (j+1) a)\over n+1})
\cos({\chi\over 2}+{\pi (j-1) a)\over n+1})}\right )
\label{eq.two.aj}
\en

\begin{figure}
\hspace{1.0cm}
\epsfxsize=12truecm
\epsfysize=12truecm
\epsfbox{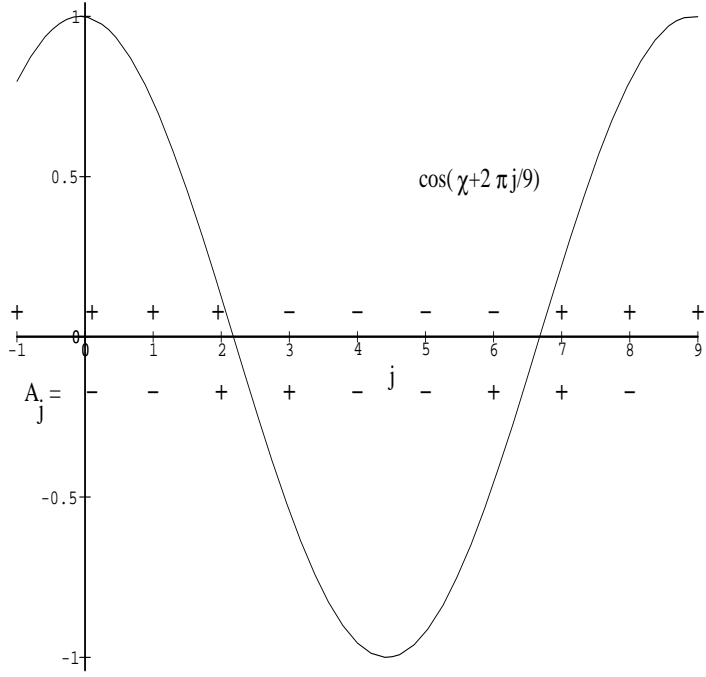}
\caption{  Boundary conditions for $N=1$, $n=8$ and $a=2$}
\end{figure}

In Figure 1. we have plotted $\cos({\chi\over 2}+{\pi j a\over n+1})$,
tabulated its sign, and tabulated the sign of the right hand side of 
(\ref{eq.two.aj}). From this we can see that the $A_j$ is generically
negative except at near a point where $\cos({\chi\over 2}+{\pi j a\over n+1})$
has a zero, where a pair of positive $A_j$ are produced. The number of 
positive pairs is related to the number of such zeroes which are in turn
related to the value of $a$. One also observes that the expression
(\ref{eq.two.aj}) is invariant under 
\eqq
\chi &\to& \chi-{\pi a \over n+1}\\
j &\to& j+1
\enn
This implements the $Z_n$ symmetry of $A^{(1)}$ Toda theories, and 
cyclically permutes the boundary conditions $A_j\to A_{j+1\;{\rm mod}\;n+1}$.

\subsection{Multi-soliton solutions}

In this section we shall generalise the two-soliton solution given above
to multi-soliton solutions with arbitrarily many constituent solitons.
Whilst this case is much more complicated, and we cannot guarantee finding
a complete set of solutions to the basic equation (\ref{eq.static.master}),
we shall give a wide class of explicit solutions.

To begin with we shall restrict our attention to multi-soliton solutions
with an even number of constituent solitons; that is, we shall assume that
the soliton corresponding to the middle node of the Dynkin diagram for $n$
odd does not feature. Thus our ansatz consists of $N$ pairs of conjugate
solitons which we shall label $p$ and $\bar{p}$ respectively. 
The real tau-function can be written in the form
\eq
1+2 \sum_{p=1}^N d_p \cos(\chi_p+{2 \pi a_p \over n+1})e^{m_{a_p} x}+\ldots
\label{eq.tau.many}
\en
where the dots compactly represent the (many) interaction terms. As in the
two-soliton case we can decompose the tau-function into sectors as in
(\ref{eq.charge.tau}). As a further simplification we shall assume that $n$ is 
large with respect to the charges of the solitons, or more particularly
that the largest charge $Q_{max}=\sum_p a_p \leq (n+1)/2$.
With this restriction, the charges in (\ref{eq.charge.tau}) take values between $Q_{max}$ and $-Q_{max}$, and only the term
\eq
\prod_{p=1}^N \left (d_p e^{i(\chi_p+{2 \pi a_p \over n+1})}\right )\left (\prod_{1\leq p <q\leq N}A^{(pq)} \right ){\rm exp}\left (\sum_{p=0}^N m_{a_p} x
\right )
\label{eq.largest.term}
\en
contributes to the charge $Q_{max}$. Substituting this into the equation
(\ref{eq.mod.stat}), we immediately find
\eq
C_\pm=\sum_{p=0}^N m_{a_p}\pm 1
\label{eq.many.energy}
\en
and the energy is as usual $-2(n+1)C_{\pm}$. Once more this pair of 
solutions are the parity inverses of one another. To determine the `positions'
of the constituent solitons we consider the terms with charge $Q_p=Q_{max}-a_p$.
Again we make the assumption the charges are chosen in such a way that only
two terms in the $\tau_j$ contribute to $T_{Q_p}$: namely the 
term containing all the solitons except the $p$-th, and the term containing
all the solitons and the $p$-th conjugate soliton. More explicitly we have
\eqq
T_{Q_p}&=&
\prod_{r\neq p} \left (D_r e^{i(\chi_r+{2 \pi a_r \over n+1})}\right )\left (\prod_{r <s:r,s\neq p}A^{(a_ra_s)} \right ){\rm exp}\left (\sum_{r\neq p} m_{a_r} x
\right )\cont
\times \left ( 1+d_p^2 A^{p\bar{p}}\prod_{r\neq p} \left( A^{(rp)}A^{(r\bar{p})}
\right)e^{2 m_{a_p} x}\right )
\label{eq.nargest.term}
\enn
Substituting this into (\ref{eq.mod.stat}) we discover that 
\eq
(d_p)^2 A^{p\bar{p}}\prod_{r\neq p} \left( A^{(rp)}A^{(r\bar{p})}
\right)={2\pm m_{a_p} \over 2\mp m_{a_p}}
\label{eq.temp1}
\en
One can prove that
\eq
A^{(rp)}A^{(r\bar{p})}={(m_{a_r}-m_{a_p})^2\over (m_{a_r}+m_{a_p})^2}
\en 
so that 
\eq
d_p=\prod_{r\neq p} {(m_{a_r}+m_{a_p})\over |m_{a_r}-m_{a_p}|}
{2\over {2\mp m_{a_p}}}
\label{eq.expression.D}
\en
Here we have made a choice for the sign of $d_p$, since we can absorb this 
sign into an $i\pi $ shift in $\chi_p$. 

Strictly speaking we have only given necessary conditions that some of 
the charge
sectors in the tau-function satisfy (\ref{eq.mod.stat}), but extensive 
numerical investigation supports the conjecture that (\ref{eq.many.energy})
and (\ref{eq.expression.D}) provide a solution to (\ref{eq.static.master}).
Moreover 
whilst we enforced fairly stringent conditions on $a_p$ to derive the necessary
constraints  on $d_p$ and $C$, many examples seem to confirm the idea
that these values give a solution even when the conditions are not met.
However, in those cases one generally might
expect other solutions not of the form (\ref{eq.many.energy}), (\ref{eq.expression.D}). 

By analogy with the two-soliton case, we should proceed by
discussing the position of the singularites 
in $\phi$. Since the solutions corresponding to $C_{\pm}$ are parity 
conjugate, and we know from the negative energy that $\phi$ is singular
somewhere on the real line $-\infty<x<\infty$, we know that one of the
two solutions must be singular in the physical region $x<0$. Unfortunately,
finding the zeroes
of the tau-function in general is very difficult. 
Numerical work suggests that it is more difficult
to find regular solutions as more solitons are introduced. As alluded to above,
solitons in the real coupling theory have negative energy, so that it seems
that the lowest energy configuration consists of adding in as many solitons
as possible without making the solution singular for $x<0$. One possibility
is that the best we can do is to
place the singularity at the wall. In this case the analysis leading 
to the expression for the total energy (\ref{eq.total.energy}) is still 
correct and yields a finite answer. Physically the infinities in the 
bulk and boundary energies exactly cancel.  

In parallel with the two-soliton case, one can ask 
which boundary condition does the solution given above correspond to; that
is for what $A_i$ does the above solution satisfy the boundary condition.
Again the answer seems to be a lot more complicated than for 
two-solitons, but nonetheless some of the features of that case do 
seem to persist. For the two-soliton case it was seen that one could
write $\tau_j(x=0)$ in the form $L W_j^2$ where all the $j$-dependence is in
$W$. From this we deduced that 
\eq
e^{\alpha^j\cdot \phi/2}|_{x=0}={|W_{j-1}W_{j+1}|\over W_j^2}
\en
and that 
\eq
A_j=\pm{\rm sign}(W_{j-1}W_{j+1})
\label{eq.A.def}
\en
where the plus/minus sign corresponds to the two solutions $C_{\pm}$.
We postulate that we can write the tau-function (\ref{eq.tau.many})
evaluated at $x=0$ in the form $L W_j^2$ where we can expand $W$ as
\eq
W_j=\sum_{\sigma_1=1,\sigma_p=\pm1} c_{\sigma_1\sigma_2\ldots\sigma_N} 
\cos(\sum_p \sigma_p \{{\chi_p\over 2}+{\pi a_p j\over n+1}\})
\label{eq.W.def}
\en
By comparing this with the tau-function with $d_p$ given in
(\ref{eq.expression.D}), we find 
\eq
c_{\sigma_1\sigma_2\ldots\sigma_N}=\prod_{r<s} \sin({\pi\over n+1}
(|\sigma_r a_r-\sigma_s a_s|))
\label{eq.c.def}
\en
and
\eq
L=4\prod_r {2\over(2\mp m_{a_r})} \prod_{r<s} {4\over {(m_{a_r}-m_{a_s})^2}}.
\en
We have checked that this ansatz for $\tau_j(0)$ is true up to six solitons.
By combining (\ref{eq.c.def},\ref{eq.W.def},\ref{eq.A.def}) we can deduce
$A_j$ for any given values of $\chi_p$, and use this to deduce a `phase' 
diagram for different boundary conditions as a function of the $\chi_p$.
One such example is given in Figure 2, where we have taken $n=8, N=2, a_1=1,
a_2=3$. In this figure we have plotted the lines in the parameter space 
$(\chi_1,\chi_2)$ along which $\tau_j(0)=0$ for $j=0,..,8$.

\begin{figure}
\hspace{1.0cm}
\epsfxsize=12.5truecm
\epsfysize=10truecm
\epsfbox{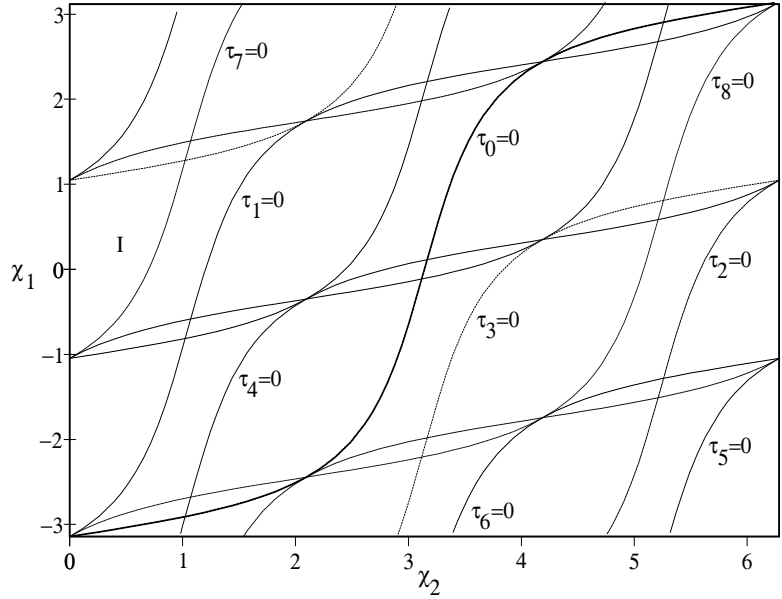}
\caption{  Boundary conditions for $N=2$, $n=8$, $a_1=1$,
$a_2=3$}
\end{figure}
 
In the region marked $I$ the boundary conditions are found to be 
\eq
(A_0,A_1,..,A_8)=(-1,1,1,-1,-1,-1,-1,1,1).
\en
The boundary conditions for the other regions
can be deduced by noting that $W_j$ changes sign as the line $\tau_j=0$ is
crossed. So by (\ref{eq.A.def}), $A_{j-1}$ and $A_{j+1}$ change sign when this
line is crossed. One can check that the resulting boundary conditions are
compatible with the $Z_n$ symmetry which is realised as
\eqq
\chi_p &\to & \chi_p -{2\pi a_p\over (n+1)}\\
j &\to& j+1
\label{eq.zn.many}
\enn
Suppose that we choose $\chi_p$ so that we lie on one of the lines $\tau_j=0$.
Then the corresponding solution for $\phi$ is singular, and in fact
\eqq
{e}^{\alpha^j \cdot \phi}&=& +\infty,\\
{e}^{\alpha^{j+1} \cdot \phi}&=& 0,\\
{e}^{\alpha^{j-1} \cdot \phi}&=& 0.
\enn
Thus along this line the coefficients $A_{j-1}$ and $A_{j+1}$ are undetermined
as we may have expected. It follows that a particularly rich variety of 
boundary conditions are allowed at the points where many such lines intersect.

If $n$ is odd we can consider static soliton solutions containing a single
constituent soliton of type $a_{N+1}=(n+1)/2$. Reality of the tau-function
constrains the associated parameter $\xi_{N+1}$ to be real, or equivalently
$\chi_p=0,\pi$. Following much the same argument as in the previous section
we initially assume that the $a_p=n+1-a_{\bar{p}}$, $p=1,..,N$ are chosen in
such a way that only one term contributes to the sector of maximal charge 
$Q_{max}=\sum_{p=1}^N a_p +a_{N+1}$ and two terms contribute to sectors of
charge $Q_p=Q_{max}-a_p$, $p=1,..,N$. This results in similar expressions
for $C_{\pm}$ and $d_p$, $p=1,..,N$ except that the sums in (\ref{eq.many.energy}), (\ref{eq.expression.D}) run from $1,..,N+1$. It 
only remains to determine the value of $d_{N+1}$. Only one term contributes
to the charge $Q_{N+1}$ sector since as there is no soliton conjugate to 
the middle soliton $p=N+1$, we cannot add a soliton-conjugate 
soliton pair to the term involving solitons of type $p=1,..,N$. Instead the 
resulting equation is
\eq
0=(C_{\pm}-\sum_{p=1}^N m_{a_p})^2-1=(m_{a_{N+1}}\pm 1)^2-1 
\en
Since $m_{a_{N+1}}=2\sin(\pi/2)=2$, we see that only the $C_-$ solution is
allowed. The value of $d_{N+1}$ is unconstrained by this or any other 
equation, so we are free to place the middle soliton anywhere! Under 
a parity transformation, the energy and the values of $d_p$, $p=1,..,N$
remain unchanged. This is in agreement with the idea that the energy of
the solution on the whole line is formally given by the sum of the half-line
energies of the solution and its parity conjugate; that is in this case 
\eq
-2(n+1)(C_-+C_-)=-2(n+1)(2\sum_{p=1}^{N+1}m_{a_p}-2)=-2(n+1)(2\sum_{p=1}^N m_{a_p}+m_{a_{N+1}})
\en
as we should have expected from the masses of the constituent solitons.

\section{Scattering solutions for $a_n^{(1)}$ Toda theories}

In the previous section we showed how one can construct static background
configurations. In this section we show 
how to 
solve the classical linearised perturbation equations around this background,
and explicitly calculate the classical scattering matrix. Once more we 
shall rely on the tau-functions given by multi-soliton solutions.

Let us assume that the field is infinitesimally perturbed about the vacuum
configuration
\eq
\phi(x,t)=\phi(x)_{vac}+\epsilon(x,t).
\en
Substituting this into the equations of motion and the boundary conditions
(\ref{eq.non.Neumann}), and keeping terms linear in $\epsilon(x,t)$ yields
\eq
\partial_{\mu}\partial^{\mu}\epsilon(x,t) +\sum_{i=0}^r n_i \alpha^i e^{\alpha^i
\cdot \phi_{vac(x)}} (\alpha^i \cdot \epsilon(x,t)) = 0
\label{eq.linear.eqm}
\en
and
\eq
\partial_x \epsilon(x,t)={1\over 2}\sum_{i=0}^r A_i \alpha^i \sqrt{n_i} e^{\alpha^i
\cdot \phi_{vac}(x)/2}(\alpha^i \cdot \epsilon(x,t)).
\label{eq.linear.bc}
\en
Far away from the boundary $x=0$, the field $\phi_{vac}(x,t)\to 0$ for finite
energy configurations, and in this limit the equation (\ref{eq.linear.eqm})
reduces to (\ref{eq.wave}) and the solution for $\epsilon(x,t)$ tends
to 
\eq
\epsilon(x,t)\to \rho_a e^{-iEt}(e^{ipx}+K_a e^{-ipx})
\label{eq.sol2}
\en
where $E^2-p^2=m_{a}^2$. This solution 
consists of the superposition of two oscillatory solutions 
corresponding to incoming and outgoing `waves' or quantum 
mechanically  `particles'. On the other hand, the field $\phi(x,t)$ corresponding to a single soliton of type $a$
has the asymptotic form
\eq
\phi(x,t)\sim  \rho_a e^{\Phi(x,t)}
\en
where $\Phi(x,t)=\sigma(x-vt)$, and $\sigma^2-\sigma^2 v^2=m_{a}^2$. Comparison of the two asymptotic behaviours suggests
that we can obtain appropriate oscillatory solutions from the soliton tau-functions if we make the identification
\eqq
\sigma&=&\pm i p\\
\sigma v &=& i E.
\label{eq.identification}
\enn

From here it is clear how to proceed. We take the tau-function $\tau_{vac}$ corresponding
to the static vacuum solution that we found in the previous section and add
in two further time-dependent solitons with the identifications (\ref{eq.identification}). If we label the incoming and outgoing solitons by
indices $I$, and $O$ respectively, then we take 
\eqq
&\sigma_I&=-\sigma_O=i p\;\;,\;\; 
\sigma_I v_I =\sigma_O v_O =i E\\
&d_I&=d_O=\epsilon\;\;,\;\;
\chi_I=-\chi_O=\psi\;\;,\;\;
a_I=a_O=b
\enn
Note that these two solitons are {\it not} a conjugate pair but are both
of species $b$. The relative phase between the two waves is given by 
\eq
K^b=e^{-2i\psi}
\en

This tau-function describes the scattering of a particle
of species $b$ on the background given by $\tau_{vac}$. As described in
section two, requiring that this tau-function satisfies one of the boundary
conditions (\ref{eq.non.Neumann}) up to $O(\epsilon)$ amounts
to the equation
\eq 
\ddot{\tau}_i-\tau_i''+2 C \tau_i'
-(C^2 -1)\tau_i|_{x=0} =0.
\label{eq.more}
\en

Considering the cases where the vacuum solution has even/odd number
of constituent solitons in turn, let us assume first that it has $2N$ solitons.
The linearity of  equation (\ref{eq.more} 
implies that we can divide $\tau_j$ into
charge sectors as in the previous equation, 
and use the equation (\ref{eq.mod.stat}) with the addition of a time-derivative
term as in(\ref{eq.more}), i.e.
\eq 
\ddot{T_k}-T_k''+2 C T_k'
-(C^2 -1)T_k|_{x=0} =0.
\label{eq.more2}
\en
There are at
least two highest charge terms in the tau-function
with charge $Q_{max}=\sum_{p=1}^N a_p +b$ of the form
\eqq
T_{Q_{max}} &=& \prod_{j=1}^N d_p \prod_{1\leq j< k\leq N} A^{(jk)}
e^{i\sum_{j=1}^N \chi_j} e^{(\sum_{j=1}^N m_{a_j})x}\cont
\times \left ( e^{i\psi}e^{iEt+ipx} \prod_{j=1}^N A^{(a_j b)}(p)+
e^{-i\psi}e^{iEt-ipx} \prod_{j=1}^N A^{(a_j b)}(-p)\right )
\enn  
where $A^{(a_j b)}(p)$ are calculated using the definition (\ref{eq.inter.def}),
(\ref{eq.identification}). Substituting this expression into (\ref{eq.more2})
we find that the scattering matrices on the two solutions corresponding to 
$C_{\pm}$ are given by
\eq
K^b_{\pm}=e^{-2i\psi}={{2ip\mp m_b^2}\over {2ip\pm m_b^2}}
{{\prod_{j=1}^N A^{(a_j b)}(p)}\over {\prod_{j=1}^N A^{(a_j b)}(-p)}}.
\label{eq.scattering}
\en
If the vacuum solution has an odd number of solitons, then we find that the 
scattering matrix is given essentially by the above expression for $K_-$, but
where the product runs from $1,..,N+1$. Note that the reflection factor
given by (\ref{eq.scattering}) only depends on the number and species
of solitons in the background solution, not on the topological charge
parameters $\chi_p$. It follows that boundary conditions related by
the $Z_n$ symmetry (\ref{eq.zn.many}) will have identical scattering 
matrices.

\subsection{Classical reflection bootstrap equations}

In the introduction, it was pointed out that in two dimensions, integrability
placed strong constraints on the $S$-matrix of the theory, and that $S$ and $K$
satisfied various algebraic constraints. In the case at hand both $S$ and $K$
are diagonal and in the classical limit $S$ tends to unity, and these two
facts ensure that most of the algebraic constraints are satisfied automatically.
However, one non-trivial check is the reflection bootstrap equation, which 
in the classical limit becomes
\eq
K^c(\theta_c)=K^a(\theta_c+i\bar{\theta}^b_{ac})
K^b(\theta_c-i\bar{\theta}^a_{bc})
\label{eq.classical.rbp}
\en
where the fusion angles are given by 
\eqq
{\theta}^b_{ac}&=&{\pi(a+b)\over 2(n+1)}\\
\bar{\theta}&=&{\pi/2}-\theta
\enn
and $a+b+c=n+1$. (For simplicity we ignore the case that $a+b+c=2(n+1)$.
For more details about fusion angles see for instance \cite{BCDS} and about
the classical reflection equation see \cite{FK1,CDR}.) 

In the equation (\ref{eq.classical.rbp}), the argument of $K^c$ is the usual 
rapidity variable $\phi$ which is related to the momentum $p$ and
energy $E$ of the incoming particle of type $a$ by
the formulae
\eqq
p&=&m_{a} \sinh(\phi)\\
E&=&m_{a} \cosh(\phi).
\enn
Note that using the identification (\ref{eq.identification}), we see that 
$\theta_p=\phi+i{\pi\over 2}$ where $\theta_p$ appears in(\ref{eq.rap.one}). 
If, as is usual, one defines a bracket notation
\eq
(x)={\sinh({\phi\over 2}+{{x i \pi}\over{2(n+1)}})\over 
\sinh({\phi\over 2}-{{x i \pi}\over{2(n+1)}})} 
\label{eq.bracket}
\en
then one can write in this notation
\eq
{A^{(a_j b)}(p)\over A^{(a_j b)}(-p)} = {{({n+1\over 2}+a_j-b)({n+1\over 2}-a_j+b)}\over {({n+1\over 2}+a_j+b)({n+1\over 2}-a_j-b)}}
\label{eq.inter.bracket}
\en
where the left hand side of this equation is one of the factors appearing
in the scattering matrix $K^b_{\pm}$ in equation (\ref{eq.scattering}).
A straightforward calculation shows that each of these factors individually
satisfy the reflection bootstrap equation. To prove that $K^b$ satisfies
(\ref{eq.classical.rbp}) it remains only to show that the remaining factor
\eq
{{2ip-m_b^2}\over {2ip+m_b^2}}=-{1\over {(n+1-a_b)(a_b)}}
\en
satisfies the bootstrap equation, and this is indeed the case. 

\section{Other simply-laced algebras}

In this section we shall make a few remarks about extending the results found
in the previous two sections to affine Toda theories based on the $D$ and $E$
series of algebras. The major technical difficulty in extending the results
to these algebras is that there is no explicit form for tau-functions which
correspond to more than two-soliton
solutions, and even those solutions that are known are considerably
more complicated than those of the $a_n^{(1)}$ theories \cite{RH}. 
Because of this, we 
shall restrict our search to finding which single soliton solutions 
\cite{ACFGZ} can be
found which satisfy the boundary conditions (\ref{eq.non.Neumann}). 
The general form for the corresponding tau-functions is 
\eq
\tau_j = 1 + \delta^{(p)}_1 d_p e^{m_p x} + .... +\delta^{(p)}_{n_j} (d_p)^{n_j}
 e^{n_j m_p x}
\label{eq.single.soliton}
\en
where $p$ labels the species of soliton as before. 
The coefficients $\delta^{(p)}_i$ have been explicitly 
calculated for all affine algebras on 
a case by case basis, and they are found to be real if and only if the 
minimal representation $\lambda_p$ of the corresponding finite Lie algebra
is real. Restricting ourselves to such real tau-functions, we substitute
the tau-functions (\ref{eq.single.soliton})
case by case into the equation (\ref{eq.static.master}) 
and solve for $C$ and $d_p$. The results are 
presented below. The labelling of nodes on the Dynkin diagram is given in
Figure 3.

\begin{figure}
\hspace{3.0cm}
\epsfxsize=12truecm
\epsfysize=18truecm
\epsfbox{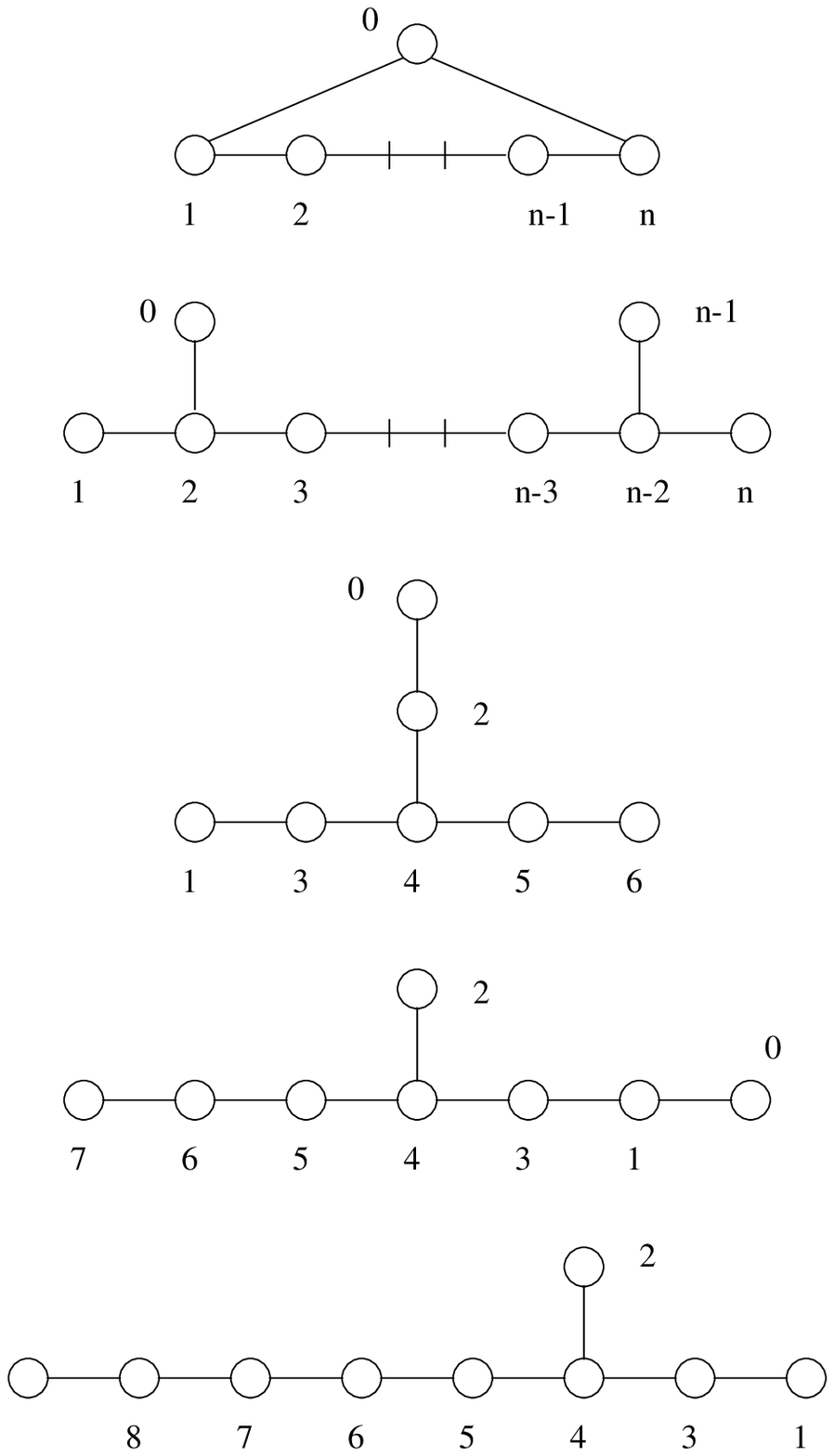}
\caption{  Root Labels for simply-laced algebras }
\end{figure}

\subsection{$d_4^{(1)}$}  

The only solution is associated with the triality invariant soliton of 
species $p=2$. The corresponding tau-functions are given by
\eq
\tau_0=\tau_1=\tau_3=\tau_4=1+d_2 e^{\sqrt{6} x},\;\;{\rm and}\;\;
\tau_2=1-4 d_2 e^{\sqrt{6} x} + (d_2)^2 e^{ 2\sqrt{6} x}
\en
and we find the solutions $C=\sqrt{6}/2,\sqrt{6}/2\pm 3\sqrt{2}/2,\sqrt{6}/2\pm
\sqrt{2}/2$ and correspondingly $d_2=-1,(5\pm 3\sqrt{3})/(-5\pm 3\sqrt{3}),
2\pm \sqrt{3}$. Only the solutions corresponding to 
$-1,(5- 3\sqrt{3})/(-5+ 3\sqrt{3}),2-\sqrt{3}$ are 
singularity free in the region $x<0$.  

For $d_2=-1$,$\tau_0\to 0$ as $x\to 0$, so that the only non vanishing 
component of ${\lambda_2\cdot \phi}\to -\infty$ in this limit. Nonetheless
the solution has finite energy, and obeys the boundary condition
$A_0=A_1=A_3=A_4=-1$ and $A_2$ is unspecified since $e^{\alpha_2 \cdot \phi}$
vanishes at the wall.
For $d_2=2-\sqrt{3}$ we find that $\tau_2$ vanishes at the wall so that 
${\lambda_2\cdot \phi}\to \infty$. We find that $A_0,A_1,A_3,A_4$ are 
unconstrained
and $A_2=1$. Finally the solution with $d_2=(5- 3\sqrt{3})/(-5+ 3\sqrt{3})$ is
regular at the wall, and obeys boundary conditions with $A_i=-1$ for all $i$.

\subsection{$d_5^{(1)}$}  

Again the only solution is associated with the species $p=2$. The corresponding
tau-functions are given by 
\eq
\tau_0=\tau_1=\tau_4=\tau_5=1+d_2 e^{2 x},\;\;{\rm and}\;\;
\tau_2=\tau_3=1-2 d_2 e^{2 x} + (d_2)^2 e^{4 x}.
\en
We find the solutions $C=(1\pm\sqrt{2\pm \sqrt{2}})$. The corresponding 
values of $d_2$ are $(2\pm \sqrt{2\pm \sqrt{2}})/(2\mp \sqrt{2\pm \sqrt{2}})$. The 
solutions which are regular in the region $x<0$ are those with
$C=(1-\sqrt{2\pm \sqrt{2}})$. These
solutions satisfies the boundary condition with $A_0=A_1=A_4=A_5=-1$ and 
$A_2=A_3=\mp 1$. 

\subsection{$d_n^{(1)}$, $n>5$}

The solitons corresponding to the tip nodes do not yield any solutions. 
The other solitons correspond to roots $\alpha_p$ with $n_p=2$ and their 
tau-functions are given by the formulae
\eqq
\tau_0=-\tau_1&=&1+d_p e^{m_p x}\\
\tau_i&=& 1+2 {\cos({(2i-1) \pi p \over 2(n-1)})\over 
\cos({ \pi p \over 2(n-1)})} d_p e^{m_p x} + (d_p)^2 e^{2 m_p x}
\;\;,2\leq i
\leq n-2 \\
\tau_{n-1}=\tau_n &=& 1+(-1)^p d_p e^{m_p x}
\label{eq.tau.dn}
\enn
The value of the coefficient involving cosine takes at least two different
values for different $i$, so inserting the difference of two 
such distinct tau-functions into (\ref{eq.static.master}) immediately
yields that $2C=m_p\pm \sqrt{2}$. Now substituting $\tau_0$ and $\tau_2$ into
the same equation we find that $m_p=\sqrt{6}$ and that $d_p=2\pm \sqrt{3}$.
Given that 
\eq
m_p=2\sqrt{2} \sin ({ \pi p \over 2(n-1)})
\en
we find that we must have $n=3x+1$, $p=2x$. With these values (\ref{eq.tau.dn})
reduces to 
\eqq
\tau_0=-\tau_1=\tau_{n-1}=\tau_n&=&1+d_p e^{\sqrt{6} x}\\ 
\tau_{3i}=\tau_{3i+1}&=&1+2 d_p e^{\sqrt{6} x}+(d_p)^2 e^{2 \sqrt{6} x}\\
\tau_{3i+2}=1-4 d_p e^{\sqrt{6} x}+(d_p)^2 e^{2\sqrt{6} x}.
\enn
Only the solution with $d_p=2-\sqrt{3}$ is non-singular in the region $x<0$.
The solution is compatible with the boundary conditions $A_{3i+2}=-1$, and the 
other $A_i$ are unconstrained.

\subsection{$e_6^{(1)}$}  

In this case only the solitons of species $p=2,4$ have real tau-functions.
Solutions which satisfy the boundary conditions can be found in both cases. 
If $p=2$, then $m_2=\sqrt{6+2\sqrt{3}}$ and solutions can be found with
$C=m_2/2,m_2/2\pm \sqrt{6-2\sqrt{3}}/2$ and $d_2=-1,5\pm 2\sqrt{6}$.
Only the solution with $d_2=5-2\sqrt{6}$ is non-singular for $x<0$, and 
satisfies boundary conditions (\ref{eq.non.Neumann}) with $A_4=-1$ and the
other $A_i=1$. 

If $p=4$, then $m_2=\sqrt{6-2\sqrt{3}}$ and solutions can be found with
$C=m_2/2,m_2/2\pm \sqrt{6+2\sqrt{3}}/2$ and $d_4=-1,5\pm 2\sqrt{6}$. The
solutions with $d_4=-1,5-\sqrt{6}$ are regular for $x<0$ and satisfy 
the boundary conditions $A_0=A_1=A_6=A_4=1$ and the other $A$'s unspecified,
and $A_i=1$ for all $i$ respectively.

\subsection{$e_7^{(1)}$}  

Only solitons of species $p=3$ and $m_3=\sqrt{6}$ can satisfy the boundary conditions with $C=m_3/2,m_3\pm \sqrt{2}/2$. The corresponding values of 
$d_3$ are $-1,2\pm \sqrt{3}$ and only the solutions with $d_3=-1,2-\sqrt{3}$
are free of singularities for $x<0$. 
The solutions with $d_3=-1$ does not seem to correspond to any sensible 
boundary condition of the type (\ref{eq.non.Neumann}) but for 
$d_3=2-\sqrt{3}$ we find that $A_1=A_2=A_6=-1$ with the other $A_i$ 
unconstrained.

\subsection{$e_8^{(1)}$}

By using the two tau-functions corresponding to $n_i=2$ and also $\tau_0$, one
can use similar arguments to those used for $d_n^{(1)},\; n>5$ to 
deduce that $m_p=\sqrt(6)$. Since there are no single solitons with such
a `mass' we find a contradiction so that there are no single-soliton solutions.

\vskip 1cm
\subsection{Scattering}

In section four we showed how by adding in two further solitons with 
imaginary momentum, we could describe classical scattering solutions on
the static `vacuum' configurations. The tau-functions had to obey the equation
\eq 
\ddot{\tau}_i-\tau_i''+2 C n_i\tau_i'
-(C^2 n_i^2 -n_i)\tau_i|_{x=0} =0.
\label{eq.more.again}
\en
As explained above, multi-soliton tau-functions for algebras other than 
$a_n^{(1)}$ are very complicated in general. However, the exception to this
rule is $\tau_0$, which has exactly the same form as for the $a_n^{(1)}$ case,
except that the interaction coefficient $A^{(pq)}(\theta)$ defined in 
(\ref{eq.inter.def}) must be generalised. For a definition of the generalisation
$X^{(pq)}(\theta)$ and a discussion of the properties it enjoys please see
reference \cite{OK}. Restricting ourselves to single soliton backgrounds,
and to linear order in the perturbation parameter $\epsilon$ we find that
\eqq
\tau_0&=&1+d_r e^{m_r x}+\epsilon e^{i \psi} e^{-iEt+ipx}
+\epsilon e^{-i \psi} e^{-iEt-ipx}+d_r\epsilon e^{i \psi} e^{-iEt+ipx+m_r x}X^{(rb)}(p)
\cont
+d_r \epsilon e^{-i \psi} e^{-iEt-ipx+m_r x}X^{(rb)}(-p).
\enn
Substituting this into (\ref{eq.more.again}) and looking at the term linear
in $\epsilon$ one recovers
\eq
e^{-2 i \psi}= _{{{\left ( (C-m_r)^2-1)((C-m_r)^2+2ip(C-mr)+m_b^2-1)
X^{(rb)}(p)-(C^2-1)(C^2+2ipC+m_b^2-1)\right )}\over {\left ( (C-m_r)^2-1)((C-m_r)^2+2ip(C-mr)+m_b^2-1)
X^{(rb)}(p)-(C^2-1)(C^2+2ipC+m_b^2-1)\right )}}}
\en

\section{Conclusions}

We have seen that integrable boundary
conditions (\ref{eq.non.Neumann}) and the equations of motion combine neatly
to yield a simple equation satisfied by
the tau-functions of affine Toda theory.
In the case of $A^{(1)}_n$ a large class of static
solutions were found, and moreover it proved relatively straightforward to 
extract the scattering data on these backgrounds. As a consistency check we
showed that the scattering matrices
satisfy the classical reflection boostrap equation. 

Still, many open 
questions remain even in this simple case. 
We only provided a plausible argument 
that the conjectured form of tau-function did indeed satisfy the boundary
condition, and we did not rule out the possibility of other solutions, amongst
which the true vacuum solution may lie . The solutions consisting 
of $2N$ solitons contained $N$ unspecified parameters $\chi_p$ 
(those with $2N+1$
solitons were specified by $N+1$ parameters). These can be thought of as
moduli for zero-modes, since the energy of the solutions is independent of 
$\chi_p$. We can change the boundary conditions satisfied by the solutions
by varying these parameters, in essentially the same way that one varies
the topological charge of solitons in the imaginary coupling theory. 
The `phase' diagram which specifies which boundary conditions
can be obtained from a particular family of solutions by varying $\chi_p$ 
seems in general very complicated and this makes it difficult to isolate
the true vacuum solution. Nonetheless, the scattering data is found to be
independent of the topological charge, so all the resulting boundary conditions
will share the same scattering matrix $K$.

A related difficulty is to find which solutions
are singularity free in the region $x<0$. The general rule seems to be that as
we add in more solitons the energy of the solution decreases rather than 
increases because of the reality of the coupling, but also the solution
develops more singularities. The trick to finding the true vacuum is 
therefore to add in as many solitons as one can without introducing 
singularities into the physical region. It may be that the best one can do 
is to have a singularity at the wall itself which is still physically 
acceptable. This may also provide a mechanism for removing the zero-modes of 
the vacuum solution, since it is conceivable that varying the parameters
$\chi_p$ would inevitably move one or more of the singularities in $\phi$
into the physical region.  

Although the formalism extends to other simply-laced algebras, the application
to these algebras is hampered by the relatively complicated 
technology for tau-functions other than those of the $a_n^{(1)}$ series.  
Whilst some partial results were obtained in these cases, it is clear that 
more powerful methods are required. Clearly one way forward is to try and
adopt the methods of \cite{OTU} which provide exact (if not explicit)
formulae for multi-soliton solutions which are equally neat for any algebra.
In their notation the tau-functions can be expressed as 
\eq
\tau_j = \langle \Lambda_j | e^{-{1\over 2} E_1(t+x)} g e^{-{1\over 2} E_{-1}(t-x)}| \Lambda_j \rangle e^{-{n_j\over 4} (t^2-x^2)}
\en
and the condition (\ref{eq.static.master}) can be written neatly as
\eq
\langle \Lambda_j | (C n_j +E_1) g (C n_j - E_{-1})| \Lambda_j \rangle=0
\en
where $g$ is an element of the group associated with the affine algebra.
It would be interesting to see if this yields a more complete solution
to the problem.

Finally, it is worth re-emphasising that analytically continued solitons
seem to play an important role in real coupling Toda theory on the 
half-line and that this provides a respectable home for them. The
model seems to be on a firmer footing than imaginary coupling theories 
with their manifest unitarity problems, and the singularities that inevitably
occur for the solitons in the real-coupling theories can be placed behind
the boundary out of harms way.


\begin{thebibliography}{10}


\bibitem{AFZ}
{A.E. Arinshtein, V.A. Fateev and A. Zamolodchikov},
\newblock{\it Quantum S-matrix of the 1+1 dimensional Toda chain}
\newblock{ Phys. Lett. {\bf B87} (1979) 389}

\bibitem{ACFGZ}
{H. Aratyn, C.P. Constantinidis, L.A. Ferreira,
J.F. Gomes and A.H. Zimerman}
\newblock{\it Hirota's solitons in affine and the conformal affine Toda models}
\newblock{Nucl. Phys. {\bf B406} (1993) 727}

\bibitem{BCDR}
{P. Bowcock, E. Corrigan, P.E. Dorey and R.H. Rietdijk}
\newblock{\it Classically integrable boundary conditions for affine Toda field theories}
\newblock{Nucl. Phys. {\bf B445} (1995) 469}

\bibitem{BCR}
{P. Bowcock, E. Corrigan and R.H. Rietdijk}
\newblock{\it Background field boundary conditions for affine Toda field theories}
\newblock{Nucl. Phys. {\bf B465} (1995) 350}

\bibitem{BCDS}
{H. Braden, E. Corrigan, P.E. Dorey and R. Sasaki}
\newblock{\it Affine Toda field theory and exact S-matrices}
\newblock{Nucl. Phys. {\bf B338} (1990) 689}

\bibitem{C}
{I.V. Cherednik}
\newblock{\it Factorizing particles on a half line and root systems}
\newblock{Theor. Math. Phys. {\bf 61} (1984) 977}

\bibitem{CM}
{P. Christe and G. Mussardo}
\newblock{\it Elastic S-matrices in (1+1) dimensions and Toda field theories}
\newblock{Int. Jour. Mod. Phys. {\bf A5} (1990) 4581}

\bibitem{CDR}
{E. Corrigan, P.E. Dorey and R.H. Rietdijk}
\newblock{\it Aspects of affine Toda theory on a half line}
\newblock{Prog. Theor. Phys. Suppl. {\bf 118} 143}

\bibitem{CDRS}
{E. Corrigan, P.E. Dorey, R.H. Rietdijk and R. Sasaki}
\newblock{\it Affine Toda field theory on a half line}
\newblock{Phys. Lett. {\bf B333} (1994) 83}

\bibitem{DGZ}
{G.W. Delius, M.T. Grisaru and D. Zanon}
\newblock{\it Exact S-matrices for non simply-laced affine Toda theories}
\newblock{Nucl. Phys. {\bf B282} (1992) 365}

\bibitem{F}
{M.D. Freeman}
\newblock{\it On the mass spectrum of affine Toda theory}
\newblock{Phys. Lett. {\bf B266} (1991) 82}

\bibitem{FLO}
{A. Fring, H.C.Liao and D.I. Olive}
\newblock{\it The mass spectrum and coupling in affine Toda theories}
\newblock{Phys. Lett. {\bf B261} (1991) 57}
 
\bibitem{FK1}
{A. Fring and R. K\"oberle}
\newblock{\it Factorized scattering in the presence of reflecting boundaries}
\newblock{Nucl. Phys. {\bf B421} (1994) 159}

\bibitem{FK2}
{A. Fring and R. K\"oberle}
\newblock{\it Affine Toda field theory in the presence of reflecting boundaries}
\newblock{Nucl. Phys {\bf B419} (1994) 647}

\bibitem{FU}
{A. Fujii}
\newblock{\it Toda Lattice Model with Boundary}
\newblock{Bonn Preprint: BONN-TH-95-18, hep-th 9510070}
 
\bibitem{GZ}
{S. Ghoshal and A. Zamolodchikov}
\newblock{\it Boundary $S$ matrix and boundary state in two-dimensional
integrable quantum
field theory}
\newblock{Int. Jour. Mod. Phys. {\bf A9} (1994) 3841}

\bibitem{RH}
{R.A.Hall}
\newblock{\it Affine Toda solitons and fusing rules}
\newblock{PhD. thesis, unpublished}

\bibitem{H}
{T.J. Hollowood}
\newblock{\it Solitons in affine Toda field theories}
\newblock{Nucl. Phys {\bf B384} (1992) 523}

\bibitem{Ki}
{J.D. Kim}
\newblock{\it Boundary reflection matrix in perturbative quantum field theory}
\newblock{Phys. Lett. {\bf B353} (1995) 213}

\bibitem{OK}
{M.A.C. Kneipp and D.I. Olive}
\newblock{\it Crossing and anti-solitons in affine Toda theories}
\newblock{Nucl. Phys {\bf B408} (1993) 565}

\bibitem{M}
{A. MacIntyre}
\newblock{\it Integrable boundary conditions for classical Sine-Gordon theory}
\newblock{Jour. Phys {\bf A28} (1995) 1089}

\bibitem{MOP}
{A. V. Mikhailov, M. A. Olshanetsky and A. M. Perelomov}
\newblock{\it Two-dimensional generalised Toda lattice}
\newblock{Comm. Math. Phys. {\bf 79} (1981) 473}

\bibitem{OTU}
{D.I. Olive, N. Turok and J.W.R. Underwood}
\newblock{\it Solitons and the energy-momentum tensor for affine Toda theory}
\newblock{Nucl. Phys. {\bf B401} (1993) 663}

\bibitem{S}
{R. Sasaki}
\newblock{\it Reflection bootstrap equations for Toda field theory}
{Kyoto preprint YITP/U-93-33; hep-th/9311027}

\bibitem{Sk}
{E.K. Sklyanin}
\newblock{\it Boundary conditions for integrable quantum systems}
\newblock{Jour. Phys. {\bf A21} (1988) 2375}


\bibitem{T}
{V.O. Tarasov}
\newblock{\it The integrable initial-value problem on a semiline:  nonlinear
\hfill\break Schr\"odinger
and Sine-Gordon equations}
\newblock{{Inverse Problems} {\bf 7} (1991) 435}

\bibitem{NW}
{N.P. Warner}
\newblock{\it Supersymmetry in boundary integrable models}
\newblock{Nucl. Phys. B450 (1995) 663}

\bibitem{ZZ}
{A.Zamolodchikov and A.Zamolodchikov}
\newblock{\it Factorized S matrices in two-dimensions as the exact solution
of certain relativistic quantum field models}
\newblock{ Ann. Phys. {\bf 120} (1979) 253}







\end{thebibliography}
\end{document}